\documentclass[final,5p,times,twocolumn]{elsarticle}
\usepackage{lipsum}
\makeatletter
\def\ps@pprintTitle{%
 \let\@oddhead\@empty
 \let\@evenhead\@empty
 \def\@oddfoot{}%
 \let\@evenfoot\@oddfoot}
\makeatother
\usepackage{amssymb}
\usepackage{amsmath}
\usepackage{multirow}
\def\tsc#1{\csdef{#1}{\textsc{\lowercase{#1}}\xspace}}
\tsc{WGM}
\tsc{QE}

\def\L{{\cal L}} 
\def\sim#1{{\widetilde{#1}}}
\def\jecl{{\sc jecl}}
\def\jec{{\sc jec}}
\def\JECL{\jecl}
\def\ssl{{\sc ssl}}
\def\AE{{\sc ae}}

\def\L{\text{{\sc {L}}}}
\def\E{\text{{\sc E}}}
\def\D{\text{{\sc D}}}
\def\MBPLS{{\sc mbpls}}
\def\mbpls{{\sc mbpls}}
\def\Val{{\sc v}}
 \def\Aro{{\sc a}}
\def\Dom{{\sc d}}
\def\Lik{{\sc l}}

\def\setup#1{{\sc setup-#1}}

\def\train{t}

\def\crossreconstruction{{\sc x-rec}}
\def\selfreconstruction{{\sc s-rec}}
\def\XREC{\crossreconstruction}
\def\SREC{\selfreconstruction}
\def\jmml{{\sc jmml}}
\def\JMML{{\sc jmml}}
\def\dcc{{\sc dcc}}
\def\cae{{\sc cae}}
\def\edcc{{\sc e-dcc}}
\def\EEG{{\sc eeg}}
\def\CCA{{\sc cca}}
\def\DEAP{{\sc deap}}
\def\RF{{\sc rf}}

\def\high{{\sc +}}
\def\low{{\sc -}}

\def\B{I\!\!B}
\def\mydef{\stackrel{def}{=}}

\def\capk{K}

\def\vX{{\textbf{X}}}
\def\cset{{\cal C}}

\def\feature{{biomarker}}
\def\features{{biomarkers}}

\begin{document}

\begin{frontmatter}

\title{Unifying EEG and Speech for Emotion Recognition: A Two-Step Joint Learning Framework for Handling Missing EEG Data During Inference}

\author[1]{Upasana Tiwari\corref{cor1}}
\ead{tiwari.upasana1@tcs.com}
\cortext[cor1]{Authors for correspondence}

\author[1]{Rupayan Chakraborty\corref{cor1}} 
\ead{rupayan.chakraborty@tcs.com}

\author[1] {Sunil Kumar Kopparapu\corref{cor1}}
\ead{sunilkumar.kopparapu@tcs.com}
\ead[url]{https://www.tcs.com}

\affiliation[1]{organization={TCS Research, Tata Consultancy Services Limited}, country={India}}

\begin{abstract}
Computer interfaces are advancing towards using multi-modalities to enable better human-computer interactions. The use of automatic emotion recognition (AER) can make the interactions natural and meaningful thereby enhancing the user experience.  Though speech is the most direct and intuitive modality for AER, it is not reliable because it can be intentionally faked by humans. On the other hand, physiological modalities like EEG, are more reliable and impossible to fake. However, use of EEG is infeasible for realistic scenarios usage because of the need for specialized recording setup.  In this paper, one of our primary aims is to ride on the reliability of the EEG modality to facilitate robust AER on the speech modality. Our approach uses both the modalities during training to reliably identify emotion at the time of inference, even in the absence of the more reliable EEG modality. We propose, a two-step joint multi-modal learning approach (JMML) that exploits both the intra- and inter- modal characteristics to construct emotion embeddings that enrich the performance of AER.  In the first step, using JEC-SSL, intra-modal learning is done independently on the individual modalities. This is followed by an inter-modal learning using the proposed extended variant of deep canonically correlated cross-modal autoencoder (E-DCC-CAE).  The approach learns the joint properties of both the modalities by mapping them into a common representation space, such that the modalities are maximally correlated. These emotion embeddings, hold properties of both the modalities there by enhancing the performance of ML classifier used for AER. Experimental results show the efficacy of the proposed approach. To best of our knowledge, this is the first attempt to combine speech and EEG with joint multi-modal learning approach for reliable AER.
\end{abstract}


\end{frontmatter}

\section{Introduction}
\label{sec:intro}

 Emotion perception in human-computer interaction (HCI) refers to the process of enabling computers to detect and respond to human emotions. With this, the HCI system can provide more personalized, adaptive, and empathetic interactions \cite{picard1995affective}. This can be achieved using a combination of inputs such as voice \cite{atmaja2022survey,de2023ongoing}, facial expressions \cite{recio2014recognizing, krumhuber2023role}, body gestures \cite{osti_10438229}, physiological signals like brainwaves \cite{kamble2023comprehensive,cui2020eeg}, heart rate \cite{agrafioti2011ecg,xiao2023spatial}, muscle movements \cite{kose2021new,lin2023review}, skin conductivity, etc. \cite{pan2023multimodal,wang2022multi,li2021multi,mittal2020m3er} to the HCI system. Speech being one of the popular modality, plays a crucial role in HCI system like customer service help-desk, healthcare, voice-based virtual assistants, chat-bots, to name a few. For instance, in call center scenario, if a user sounds low and unhappy, a virtual assistant might offer more empathetic support or simplify the instructions. In mental health, AI-driven emotion recognition tools help therapists monitor patients’ emotional well-being. Despite its potential, the challenges like data scarcity, data privacy, security, cultural bias, ethical concerns, etc. are at the forefront in ongoing research in this area. As HCI continues to evolve, the integration of multi-modal information in emotion recognition aims to make technology more human-centric, fostering more natural and meaningful interactions \cite{shimojo2001sensory}.

 Over the past decade, a majority of automatic emotion recognition (AER) studies have been focused on uni-modal emotion recognition using only single modality \cite{sajjad2020clustering,aftab2022light,yao2020speech}. However, human emotions being a complex representation of individuals physiological and psychological state cannot be reliably captured with uni-modal signals, specially non-physiological signal (e.g speech and facial expression). 
Voice, though popular, is not always a reliable modality for identifying human emotion due to several reasons, (a) emotion in voice varies significantly across individuals, cultures, and languages making them ambiguous, (b) background ambient noise can distort the voice signals, (c) multiple emotions are often blended together, and (d) emotions can be subtle, as the speakers can deliberately hide their emotions. However, even so, facial expressions and speech are still the dominant and widely used external channels for expressing emotion. The
study in \cite{mehrabian2017communication} showed that these two modalities account for 93\%
of the emotional information in human communication, thus making them critical for AER.
For these reasons, emotion recognition systems combine voice with other reliable modalities like physiological data, to achieve more accurate and reliable understanding of speakers' emotion. Speech based emotion recognition relies on external cues like pitch, tone, and rhythm, which can be influenced by environmental noise, cultural differences, and a person’s conscious control over their voice. In contrast, brainwaves captured through \EEG\ directly measures electrical brain activity, providing a more internal representation of emotional states \cite{chanel2006emotion,coan2004frontal}. Since emotions originate in the brain, \EEG\ signals offer deeper and more direct insight into emotional states, even when they are not outwardly expressed \cite{kamble2023comprehensive,cui2020eeg}. 

 The multi-modal approach addresses the limitations of using speech as a lone modality for AER. For instance, while a person might mask their emotions in their voice, physiological signals like EEG can still capture the underlying neural patterns of stress, anxiety, or excitement. Additionally, EEG is less affected by language, accent, or cultural differences, making it more universally applicable. By fusing data from both modalities, machine learning models can identify correlations between brain activity and vocal features, improving classification accuracy and reducing misinterpretations. The combined information from these two independent modalities can 
also enhances system reliability, especially in noisy or unpredictable environments. This approach based on multi-modality
 can be used in applications like mental health monitoring, personalized learning, where %
 high accuracy emotion recognition is crucial. 
 However, EEG requires specialized hardware (e.g. EEG headsets) and is more intrusive, which poses challenges for large-scale, everyday use. Despite these challenges, the synergy between speech and EEG offers a promising path toward more effective and empathetic human-computer interactions \cite{abhang2015correlation}. 
  Speech being hands-free as well as easy to use, and EEG being reliable, here in this paper one of our primary objective is to make use of EEG brainwaves while learning better embedding space of each emotion in audio modality, and facilitate missing modality (e.g. EEG) testing.
 This way, one can take advantages from reliable modalities like EEG at the time of audio+EEG model training, and reliably detect emotion at the time of testing, even if a more complex modality like EEG is missing.      

Though there has been numerous work on EEG based multi-modal emotion recognition (EMER) \cite{liu2024eeg}, either focusing on peripheral physiological signals \cite{zhang2020emotion,sharma2021multimodal,luo2022semi,liu2021comparing} or facial expressions as well as the eye movements \cite{li2020eeg,huang2017fusion,zhu2020valence,wang2023multimodal,wu2022investigating,tang2017multimodal}, but direct use of the complementarity and redundancy between
EEG and speech for emotion recognition has not been well explored \cite{subramanian2021multimodal,pan2023multimodal,wang2022multi,li2021multi,ghoniem2019multi}.
The work in \cite{subramanian2021multimodal} employ various feature-level fusion on three modalities, namely facial images (FER and Ck+ datasets), speech (RAVDESS dataset), and EEG (SEED-IV dataset), respectively, for multi-modal AER. In \cite{pan2023multimodal}, the authors proposed three different deep learning systems, GhostNet, a lightweight fully convolutional neural network (LFCNN) and a tree-like LSTM (tLSTM) for feature extractions from facial expression, speech and EEG modalities, respectively. Finally, decision-level fusion strategy was adopted to integrate the recognition results of the above three modalities. In \cite{li2021multi}, the authors used CNN to extract EEG features and bidirectional long
short term memory (BiLSTM) neural networks to extract the speech
features. Furthermore, these embeddings were combined using feature-level fusion followed by a softmax layer to recognize the levels in arousal and valence dimension. They used EEG signals from DEAP dataset for emotion recognition. The audio corresponding to the video stimuli has been used as other modality.
The work in \cite{wang2022multi} developed four different baseline systems, namely, Identification-vector +
Probabilistic Linear Discriminant Analysis (I-vector + PLDA), Temporal Convolutional Network (TCN), Extreme
Learning Machine (ELM) and Multi-Layer Perception Network (MLP). Furthermore, two fusion strategies on
feature-level and decision-level respectively were employed to combine EEG and speech modality. The work was conducted on the private multi-modal dataset, MED4, having recording environment with natural noises and an anechoic chamber. Result shows that fusion methods also enhance the robustness of AER in the noisy environment.
The work in \cite{ghoniem2019multi} employed hybrid fuzzy-evolutionary computation methodologies for both uni-modal EEG and Speech data, followed by decision-level fusion to perform multi-modal emotion recognition. The work used SAVEE and
MAHNOB dataset for Speech and EEG, respectively.

Most of these existing EMER work combine EEG and other complementary modality either with early or late fusion strategies. These traditional fusion method has limitations like it cannot guarantee the integrity and representativeness of fusion information simultaneously.
Also, there are no studies so far that have endeavored to learn speech and EEG simultaneously via coordinated learning for recognizing
human emotion. 
This has motivated us to propose a joint multi-modal learning approach
that fuses speech and EEG modalities in a coordinated subspace to learn the conflicting and complementary information simultaneously for robust and reliable AER.
In this work, we jointly learn the corresponding involuntary internal brain response and the voluntary external behavior of human emotion, i.e. via EEG and speech, respectively, which we believe simultaneously fuses the complementary information from both the modalities for reliable emotion recognition.

\begin{figure}[t]
    \centering
    \includegraphics[width=0.7\linewidth]{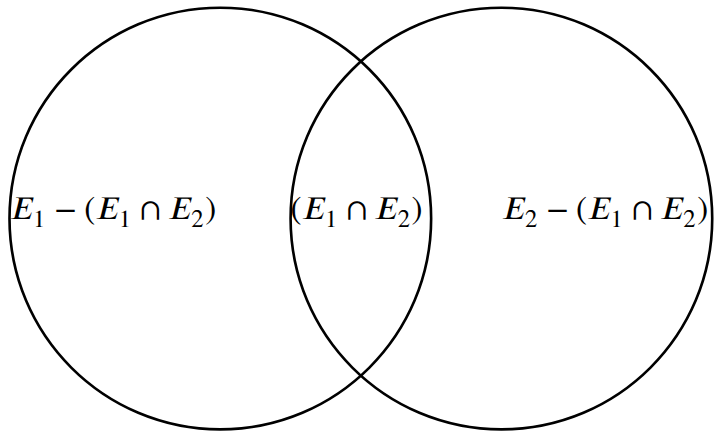}
    \caption{Two class emotion representation ($E_1$ and $E_2$) with shared characteristics ($E_1 \cap E_2$). 
    }
    \label{fig:emotion_class}
\end{figure}

\begin{figure}[t]
    \centering
    \includegraphics[width=0.9\linewidth]{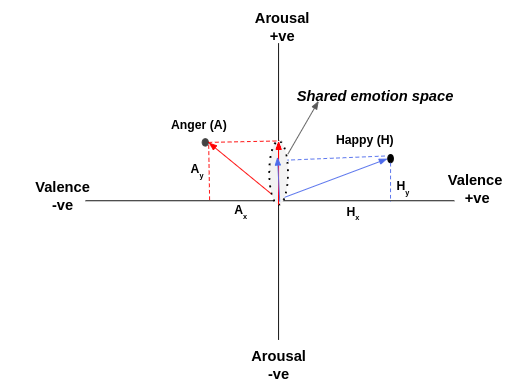}
    \caption{Sample representation of the shared characteristics ($H_y$, $A_y$) and class specific characteristics ($H_x$, $A_x$) for two emotion classes, Anger ($A$) and Happy ($H$) in the valence-arousal emotion space.}
    \label{fig:2d_emo_model}
\end{figure}

In addition, existing multi-modal fusion strategies do not fully
focus on the fusion of the inter-modal information on top of intra-modal fusion.
The intra-modal fusion is motivated from the fact that no emotion can exist in reality as a single affective (or psychological) state. Instead, human emotions are comprised of a collectively related emotional states which are result of the variations on a shared emotion content \cite{ekman}. 
Thus to enrich the representative biomarkers for efficient 
emotion recognition, they must not only capture the characteristics specific to a particular emotion,  but should also capture the common characteristics shared among all the emotion classes. 
{Figure \ref{fig:emotion_class}, for the purpose of representation, shows two emotion classes $E_1$ and $E_2$. Let $E_1 \cap E_2$ represent the characteristics 
common
to $E_1$ and $E_2$. In order to discriminate $E_1$ and $E_2$, most researchers in literature concentrate on extracting biomarkers which in some form try to capture $E_1 - (E_1 \cap E_2)$ and $E_2 - (E_1 \cap E_2)$ so that $E_1$ and $E_2$ can be distinguished without ambiguity. 
However, 
there is a %
biomarker space, $(E_1 \cap E_2)$, common to both $E_1$ and $E_2$ which is not utilized. In this paper, we try to embark on a process that not only captures  $E_1 - (E_1 \cap E_2)$ and $E_2 - (E_1 \cap E_2)$ but also utilizes the characteristics shared by $E_1$ and $E_2$, namely, $E_1 \cap E_2$. We call this joint emotion class learning, where we not only learn  characteristics specific 
to a class
but also learn characteristics that are common across the classes.}
For example, %
two emotions, namely, $A$ ({\em Anger}) and $H$ ({\em Happy}) are shown as a vector in the arousal-valence space
in Figure \ref{fig:2d_emo_model}.  
These emotion vectors can be decomposed into the valence ($H_x$, $A_x$) and the arousal ($H_y$, $A_y$) axis. It can be clearly observed that the two emotions have shared characteristics, equivalent of $(E_1 \cap E_2)$ along the arousal axis ($H_y$ and $A_y$ overlap) while having emotion specific characteristics along the valence axis, namely, $H_x$ and $A_x$.

In this paper, we devise a two step joint multi-modal learning pipeline that exploits the intra- as well as inter- modal characteristics to enrich the input modalities towards improved and reliable AER. We term this approach as \jmml.
The approach is designed to first do a joint intra-modal learning in the respective input modalities by employing {\jec-\ssl \cite{tiwari2024joint},} followed by joint inter-modal learning using proposed extended variant of deep canonically correlated cross-modal autoencoder. We term this variant as \edcc-\cae. In joint intra-modal learning, \jec-\ssl\ consists of two components, namely Joint Emotion Class Learning (\jecl) followed by Self-similarity Learning (\ssl). \jecl\ simultaneously captures both the class-specific %
 (intra-)  and class-shared (inter-) characteristics while 
 \ssl\ decomposes the  jointly learnt emotion embedding into a common latent structure such that it ensure the maximal covariance with the original %
 input signal while retaining %
 the joint characteristics. 
 Furthermore, 
 intra-modal
\jec-\ssl\ embeddings
 are fed into \edcc-\cae\ to perform joint inter-modal learning, where the system jointly learns both the modalities by mapping them into a common representation space, such that the modalities are maximally correlated.
A type of multi-modal learning, where the learnt representation of respective input modalities exist in their own space but are coordinated through a
correlated subspace through a similarity (e.g., Euclidean distance) or structure constraint (e.g., partial order) is also called coordinated learning in 
some
literature and is distinct from joint learning \cite{baltruvsaitis2018multimodal}. However we use the notation of joint learning (\jmml) to represent the coordinated multi-modal learning proposed in this work.
In the proposed work, we make use of bimodal non-parallel Speech and EEG data where our main aim is to enrich the most
direct and intuitive communication media in human interactions and widely used modality for automatic emotion recognition (AER), i.e Speech, by complementing it with one of the most reliable physiological modality, EEG, through our proposed \jmml\ approach. Given the complexity of recording EEG data which limits it's availability in realistic scenarios, our system make use of the EEG only during training to enrich the Speech representation. Thus, 
\jmml\ facilitates the Speech representation enhancement with embedded EEG learning without the need of EEG availability at the test time, making it more realistic for the real-world scenarios.
Lastly, the final \jmml\ based joint modal embeddings are learnt using final-stage classifier to perform EEG emotion detection. Experimental results shows the efficacy of our proposed approach. To best of our knowledge, this study is the first attempt to
combine speech and EEG with joint multi-modal learning approach for reliable AER 
The rest of the paper is organized as follows: In Section \ref{sec:method} we describe the methodology which is the main contribution of this paper. In Section \ref{sec:exp} we use speech and EEG as two modalities and demonstrate the efficacy of the proposed pipeline and discuss the experimental results in Section \ref{sec:results} and conclude in Section \ref{sec:conclu}. We would like to mention that  this work is an extension of \cite{tiwari2024joint} where we extend the \jec-\ssl\ framework for multi-modal inputs along with experimental results. 

\section{Methodology}
\label{sec:method}

\begin{figure}[t]
	\centering
        \includegraphics[width=0.9\linewidth]{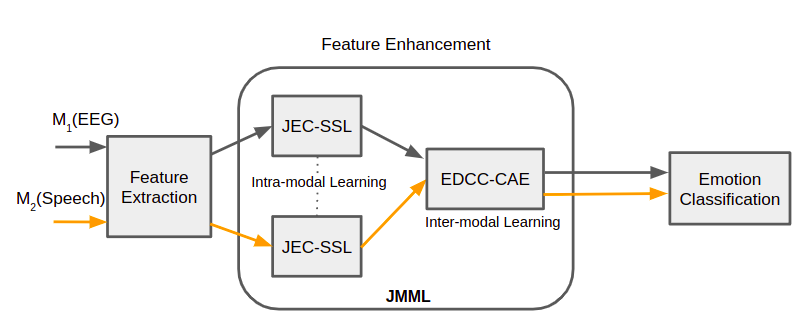}
        \caption{System Overview.}%
        \label{fig:system}
\end{figure}

The proposed approach 
adopts 
a two-step \textit{joint} multi-modal learning to enrich the speech representation by infusing it with EEG based complementary 
representation 
for improved SER.
As a first step, it
performs %
a 
\textit{joint} intra-modal learning (\jec-\ssl) which is then followed by \textit{joint} multi-modal learning (\jmml) as the second step. 
A %
system 
using 
two modalities
is shown in Figure \ref{fig:system}, 
a feature extraction module extracts
(1) EEG biomarkers %
from the raw EEG signal, and (2) Acoustic biomarkers
from speech signal, separately.
As a first step, the
\jec-\ssl\ block, 
acting on each modality separately, 
performs joint intra-modal learning.
At this stage, 
in addition to 
the individual emotion class characteristics (example, $H_x$ and $A_x$ in Figure \ref{fig:2d_emo_model}), 
characteristics that are common across the different classes (example, $H_y$ and $A_y$ in Figure \ref{fig:2d_emo_model}) 
are learnt 
\textit{jointly}.
In the second step,
these \jec-\ssl\ embedding, one for each modality,
are fed into 
EDCC-CAE block
which 
jointly learns both the modalities by mapping them into a common representation space, such that %
the modalities are maximally correlated.
\subsection{Feature Extraction}
\label{sec:features}

We describe in brief the state-of-the-art biomarkers extracted from each of the two modalities.

\noindent {\bf EEG Biomarkers:} We 
extract
state-of-the-art temporal and spectral EEG biomarkers \cite{bao2011pyeeg}.
In the temporal domain, we extract four types of \features,
namely, (a) Fractal dimension using Higuchi (HFD) and Petrosian algorithm (PFD), (b) Hjorth mobility and complexity parameters, (c) Detrended Fluctuation Analysis (DFA), and (d) Hurst Exponent. 
We extract %
(a) Power Spectral Intensity (PSI), (b) Relative Intensity Ratio (RIR) and (c) Spectral Entropy, are extracted from standard EEG frequency bands, namely, $\theta$ (4 - 8 Hz), $\alpha_{low}$ (8 - 10 Hz), $\alpha_{high}$ (10 - 13 Hz), $\beta$ (13 - 25 Hz) and $\gamma$ (25 - 40 Hz) 
as spectral features.

 \begin{figure}[t]
 	\centering
    
    \includegraphics[width=0.9\linewidth]{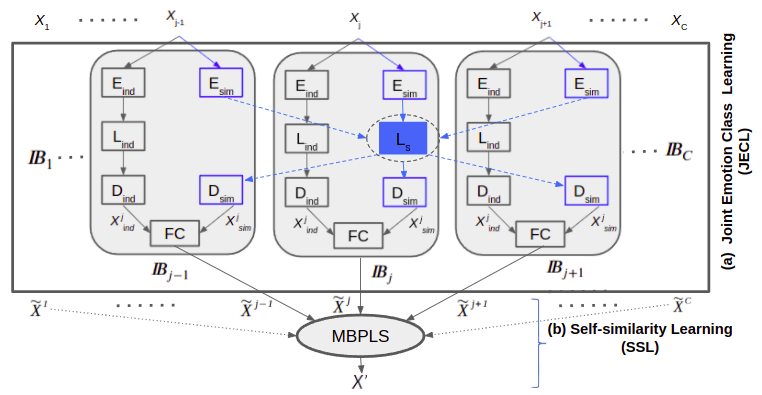}
    \caption{Joint Emotion Class Learning (\JECL) for $C$ classes, followed by \mbpls\ based \ssl (\jec-\ssl). The $j^{th}$ class learns from all other classes through its similarity latent space as well as its independent latent space.} 
    \label{fig:jec-ssl}
 \end{figure}

\noindent {\bf Acoustic Biomarkers:} We extract 
$88$ %
acoustic features %
from each audio file with extended Geneva Minimalistic Acoustic Parameter Set (eGeMAPS)  using eGeMAPSv01a \cite{eyben2015geneva}
configuration file of 
the
OpenSMILE toolkit \cite{opensmile}. 
There are 
a total of 
$18$ acoustic features, namely
  Pitch, Jitter, Shimmer, formant related energy, MFCCs 
  also known as  
low-level descriptors (LLDs)
and  high-level descriptors (HLDs) are computed (mean, standard deviation, skewness, kurtosis, extremes, linear regressions, etc.) for each of the LLDs. 
    \subsection{Joint Intra-modal Learning (\jec-\ssl\ \cite{tiwari2024joint})}
    The system performs intra-modal feature enhancement 
    within both the modalities, %
    individually.
    \jec-\ssl \cite{tiwari2024joint} mechanism is adopted to enhance the EEG and speech representation, individually, through %
    a subspace mapping. %
    The learnt embeddings, thus not only capture the 
    characteristics of different emotions, but also 
    capture 
    the similarities 
    across %
    different 
    emotion classes.

    \subsubsection{Joint Emotion Class Learning (\jecl)}
	\label{sec:jecl}
 Let $e(t) \stackrel{\Delta}{=}\{e_j(t)\}_{j=1}^E$ represent the raw \EEG\ signal consisting of $E$ channels and let $a(t)$ represent the acoustic or speech signal. Note that in our formulation we do not require $e(t)$ and $a(t)$ to be parallel, namely data collected %
 for 
 the same subject %
 performing 
 the same activity. In fact, in all our experiments 
 $e(t)$ and $a(t)$ are not related; 
 we 
 only
 assume knowledge of 
 emotion labels corresponding to $e(t)$ and $a(t)$. Let %
 $\vX$ represent 
 the 
 feature extracted from the raw signal (either $e(t)$ or $a(t)$). 
 
The aim of \JECL\ is to transform the input 
 $\vX$ into an embedded space %
 such that 
 the 
 resulting 
 embedding not only captures the class-specific properties 
 (intra-class)
 but also the 
 properties
 shared across all the emotion classes %
 (inter-class).
	As shown in Figure \ref{fig:jec-ssl}, %
 $\{1, \cdots, (j-1), j, (j+1), \cdots, C\} = \cset$ are the 
 $C$ emotion classes  
 and 
 $\vX^j$ represents %
 a sample belonging to the emotion class $j$. 
 To facilitate %
 \jecl, for %
 emotion 
 class $j \in \cset$,  %
  we learn,
 (a) an {\em independent branch} function $g^{j}_{ind}$  corresponding to the $j^{th}$ emotion class
 \begin{eqnarray}
     \vX^{j}_{ind} &=& g^{j}_{ind}\left (\vX^j;\theta^j_{ind}\right )
     \label{eq:ind_sim_ind}
 \end{eqnarray}
 where $g^{j}_{ind}$, parameterised by 
	$\theta^j_{ind}$, captures the $j^{th}$ intra-class characteristics (black arrow path in the $\vX^j$ block, Figure \ref{fig:jec-ssl}) 
 and 
 (b) a {\em similarity branch} function $g^{j}_{sim}$ corresponding to %
 all the $C$ emotion classes
 \begin{eqnarray}
     \vX^{j}_{sim} &=& g^{j}_{sim}\left (\vX^j;\theta^j_{sim},\bar \theta^j_{sim} \right )
     \label{eq:ind_sim_sim}
 \end{eqnarray}
    where, $g^{j}_{sim}$, parameterised by $\theta^j_{sim}$ and  $\bar\theta^j_{sim}$, captures the inter-class similarities (blue arrow path in the $\vX^j$ block, Figure \ref{fig:jec-ssl})  %
 shared across %
 all the 
 $C$
 emotion classes.
	These two branches, learnt %
 for every 
 emotion class $j \in \cset$ %
 collectively forms an {\em Emotion Block} $\B$ (see Figure \ref{fig:jec-ssl}).
	We implemented both these functions (\ref{eq:ind_sim_ind}) and (\ref{eq:ind_sim_sim}) in %
 {$\B$}
 as an autoencoder (\AE) consisting of an encoder ($E_{ind}, E_{sim}$), a latent ($L_{ind}, L_{sim}$) and a decoder ($D_{ind}, D_{sim}$) layers.
 Specifically, as shown in Figure \ref{fig:jec-ssl} the shared parameter $\bar\theta^j_{sim}$ in the latent space, $L_{sim}$, of the similarity branch is shared across all the 
 $C$
 emotion classes. %
	The objective is to minimise the reconstruction loss 
{$l_r^j()$} %
 which  is the sum of two measures, namely, (a) the cosine similarity and  (b) the Kullback–Leibler divergence (KLD). While the cosine similarity %
 maximizes the proximity between the predicted and the target vector %
 (enables the network to learn the variational mapping of the input) the KLD 
 controls the sample divergence from the {centroid} of %
 class $j$. Together this ensures %
 that 
 the learnt joint embedded space to be %
 discriminative while also being %
 closely packed.
 Specifically, we minimise the cost function,
 	\begin{equation}
	\Theta^{*}_{ind},\Theta^{*}_{sim},\bar \Theta^{*}_{sim} 
 = 
 \min\limits_{\theta^{1,2,\cdots,C}_{ind},\theta^{1,2,\cdots,C}_{sim},\bar \theta^{1,2,\cdots,C}_{sim}} 
 \left \{ \sum_{j=1}^{C} l_r^j \left (\vX^j,\vX^{j}_{ind}, \vX^{j}_{sim} \right ) \right \} %
 \label{eq:cost_mininization}
	\end{equation}
 where $\Theta^*_{ind},\Theta^{*}_{sim},\bar \Theta^{*}_{sim}$ 
 represents the set of all the learnable parameters 
 $$%
 \overbrace{
 \theta^{*1}_{ind}, \theta^{*2}_{ind}, \cdots, \theta^{*C}_{ind}}^{\Theta^*_{ind}},
 \overbrace{\theta^{*1}_{sim}, \theta^{*2}_{sim}, \cdots, 
 \theta^{*C}_{sim}}^{\Theta^{*}_{sim}},
 \overbrace{\bar\theta^{*1}_{sim}, \bar\theta^{*2}_{sim}, \cdots, \bar\theta^{*C}_{sim}}^{\bar \Theta^{*}_{sim}}
 $$ which together minimise (\ref{eq:cost_mininization}). 
	Note that 
	$ %
\vX_{ind}^{j}$ and  %
	$%
\vX_{sim}^{j}$ %
in (\ref{eq:cost_mininization}) are the 
intra-class and 
inter-class 
 reconstructed mapping 
 for the emotion class $j$ respectively. %
 Furthermore, we stack %
 inter and intra %
 emotion class representations to obtain a joint emotion class representation, namely, 
	\begin{equation}
	    \sim{\vX}^j = \left \{\vX_{ind}^j,\vX_{sim}^j \right \}.
     \label{eq:joint-embedding}
	\end{equation}

 {We hypothesize that $\sim{\vX}^j$ is a better representation of the emotion than $\vX^{j}$ itself.}
    The \AE\ branch %
 in each $\B$ aims to reconstruct %
 itself %
 such that jointly learnt embeddings ($\sim{\vX}^j$) of a class $j$ are 
 discriminative and yet
 mapped %
 in the close %
 proximity of %
 $j$.
	\subsubsection{Self-similarity Learning (\ssl)}
	\label{sec:mbpls}
 For every 
 emotion class 
 $j \in \cset$, %
 the emotion block $\B_j$ of \jecl\ learns 
 the joint embedding 
 (\ref{eq:joint-embedding})
(see 
Figure \ref{fig:jec-ssl}).  The task %
 is to combine all of these jointly learnt $\left \{\sim{\vX}^j\right \}_{j=1}^C$ %
 so that %
 the similarities shared %
 across all 
 $\B_j$'s is captured while retaining %
 the distinctive characteristics of each individual class. %
 We adopt self-similarity learning (\ssl) by projecting these 
 $\left \{\sim{\vX}^j\right \}_{j=1}^C$
 to a discriminative common subspace, such that it has %
 maximal covariance with the original %
 sample $\left \{\vX^j \right \}_{j=1}^C$.

\noindent {\bf Multiblock Partial Least Square (\MBPLS)}:
 We use %
Multiblock Partial Least Square (\MBPLS) %
 \cite{baum2019multiblock}, %
 an extension of Partial Least Square (PLS), 
 to perform the \ssl\ %
 of each $\B$. 
  \MBPLS\ takes the \jecl\ based joint embeddings, namely  
  $\left \{\sim{\vX}^j\right \}_{j=1}^C$,
  as input 
and aims to find 
a suitable subspace projection which 
maximizes the covariance between the projected input and the original sample $\vX$ (target variable).
	Instead of learning an interpretative model for the entire data matrix $\sim{\vX} \mydef \left [\sim{\vX}^1, \cdots, \sim{\vX}^C \right ]$ obtained by simple concatenation of each $\sim{\vX}^j$, \MBPLS\ learns model parameters for each individual data block $\sim{\vX}^j$. This makes \MBPLS\ capable of capturing the
 relative importance measure, i.e.  how much each emotion block $\B_j$ contributes to the prediction of target 
 $\vX$.

	The general underlying model of %
 PLS is %
 an iterative process %
 to %
 identify 
 $K$ Latent Variable (LV). %
For every $k^{th}$ LV %
($k = 1, \cdots, K$), \MBPLS\ aims to find {\em loading vectors} $p_k$ and $v_k$ which 
project the data
to LV scores $t_{sk}$ and $u_k$, indicating maximal covariance.
Subsequently, the explained variance is {\em deflated} to extract more %
LV's. %
Deflation for the $k^{th}$ LV is calculated as:
	\begin{equation}
	\sim \vX_{k+1}^j=\sim \vX_{k}^j - t_{sk}p^{T}_{k}
    \label{eq:5}
	\end{equation}
\MBPLS\ assigns an importance score ($i_{jk}$) to each $\sim{\vX}_{k}^j$ in the prediction of $\vX$. 
After the completion of an iterative process, we get 
{\em loading matrices} 
$P = [p_1|\cdots|p_\capk]$ and  $V = [v_1 | \cdots | v_\capk]$; projections of $\sim{\vX}$ represented as $T_s = [t_{s1} | \cdots | t_{s\capk}]$ (also known as the {\em factor matrix}); and projection of response variable $\vX$ represented as $U = [u_1 | \cdots | u_\capk]$. 
 So, for a given set of input $\sim{\vX} = \{\sim{\vX}^j\}_{j=1}^C$ received from $\{\B_j\}_{j=1}^C$, the \MBPLS\ based projection of each input block $\sim{\vX}^j$ and the target $\vX$ can be expressed as
	\[ \sim{\vX}^j =T_{s}{P}_j^{T}+{E_j},\] 
        \[ \vX = UV^{T}+E_X,\]  
        \[ \vX = \sim{X}\beta+E, \]  
	where, 
$E_j$ and $E_X$ are the error terms \cite{mbpls-python}. 

 Finally, the \ssl\ %
 function ($\phi_{ss}$) is implemented via \MBPLS\ based projections of jointly learnt emotion blocks  to a discriminative common latent structure. Mathematically, %
	\begin{equation}
	\text{\MBPLS}\left ( \left [\sim{\vX}^{1},\cdots,\sim{\vX}^{C} \right ],\vX \right )\xrightarrow{\phi_{ss}}  \vX' 
    \label{eq:6}
	\end{equation}
	where, $\vX'$ is the \MBPLS\ based prediction of input $\sim{\vX}$ corresponding to target $\vX$; obtained via discriminative subspace learning of jointly learnt emotion blocks using \jecl\ followed by \MBPLS\ (see Figure \ref{fig:jec-ssl}).
 Thus,   \jec-\ssl\ outputs a joint embedding $\vX'$ which captures 
 both
 discriminative (intra) and joint (inter) emotion class characteristics. 
 Finally, we train a final-stage classifier to perform emotion classification on these embeddings.

\subsection{Joint Multi-Modal Learning (\jmml)}
\label{sec:jmml}

Let $M$ %
be the total number of modalities and let $\vX_m$ represent the sample belonging to the modality $m$.
For every 
modality 
$m$, %
\jec-\ssl\ learns the intra-modal embedding, $\vX'_{m}$ as shown in Figure \ref{fig:jec-ssl}.  
\jmml\ aims to
learn
correspondence between different %
modalities such that %
the %
multi-modal representation incorporates %
complementary information from %
the input modalities. 

\begin{figure}[t]
	\centering
	\includegraphics[width=0.9\linewidth]{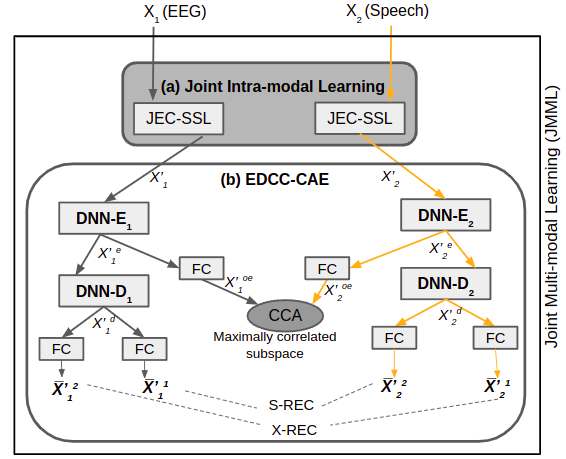}
	\caption{{Joint Multi-modal Learning (\JMML) for bimodal data; \selfreconstruction\ and \crossreconstruction\ represents the reconstructed version of self and other respective modality}}
	\label{fig:jmml}
\end{figure}

We implement an extension of deep canonically correlated cross-modal autoencoder, \dcc-\cae\ 
\cite{dumpala2019audio},
to %
map the modalities onto a common representation space, such that %
the 
modalities are maximally correlated. 
To the best of our knowledge, this work is the first attempt to 
jointly learn
\EEG\ and speech %
modalities for emotion recognition using 
extended variant of
\dcc-\cae.
Note that 
the performance of \dcc-\cae\ is 
not affected 
in the absence
of an input modality \cite{dumpala2019audio}, namely only \EEG\ or only speech, thereby
addressing the missing modality scenario during inference. 
However, knowledge of all the modalities 
is assumed to be known during training, 
this can be  %
a serious drawback in using \dcc-\cae.

To overcome this limitation, we extend %
\dcc-\cae\ network 
to not only 
self-reconstruct 
(\selfreconstruction) itself but also %
cross-reconstruct (\crossreconstruction) the other modality. This way, %
the network training process 
is %
not explicitly dependent on prior knowledge of the 
missing modality
during inference.
We term this variant of \dcc-\cae\ as \edcc-\cae\ (extended \dcc-\cae) and %
is implemented as an %
\AE\ that takes intra-modal (\jec-\ssl) embedding as an input and learns the multi-modal embedding such that the canonical correlation between the transformed representations is maximized. Namely, 
\begin{eqnarray}
    {\vX'}_{m}^e & = & f_m^e \left (\vX'_m ; \theta^e_m \right ), \nonumber \\
    {\vX'}_{m}^d & = & f_m^d \left ({\vX'}^e_m ; \theta^d_m \right ), \forall\ m = 1, \cdots, M
    \label{eq:jmml-functions}
\end{eqnarray}   %
where 
$f_m^e$ and $f_m^d$  denote the non-linear transformations of the encoders 
and decoders, 
respectively for the modality $m$. Figure \ref{fig:jmml} represents the \jmml, where, DNN-E$_{m}$ and DNN\hbox{-}D$_{m}$ represent the encoders and decoders, respectively for the modality $m$.
The weight matrices
of  DNN-$E_{m}$ and DNN-D$_{m}$, are $\theta^e_m$ and $\theta^d_m$ respectively.
And the objective function to be minimised is %
 \begin{equation} 
 \min_{\theta_{1}, \cdots,\theta_{Z}} -\mbox{\CCA} \left ({\vX'}_{1}^{oe}, \cdots, {\vX'}_{M}^{oe} \right )+ \mbox{\SREC} + \mbox{\XREC}
 \label{eq:my_jmml_loss}
 \end{equation}
 where,
    \[ \mbox{\SREC}= \sum_{m=1}^M L_r({\vX'}_{m}^d,{\bar \vX}_{m}^{'d}),\] and 
    \[ \mbox{\XREC}= \sum_{m =1}^M \sum_{n \ne m =1}^M L_r({\vX'}_{m}^d,{\bar \vX}_{n}^{'d})\] %
and, $L_r$ %
is the binary cross-entropy loss function. 

As can be seen in Figure \ref{fig:jmml} %
each DNN-$D_m$, %
takes the encoded version of a %
modality, $m$, as the input, and
reconstructs %
itself (\selfreconstruction) as well as %
other modalities (\crossreconstruction).
Finally, for every modality $m$, DNN-$D_m$ reconstructs 
${\bar \vX}_m^{'m}$ (self) 
and ${\bar \vX}_{m}^{'n}$ (cross) from 
the input %
$\vX_m$,  $\forall m,n \in \{1, \cdots, M\}$, and $m \neq n$. %

\section{Experimental Setup}
	\label{sec:exp}
    \subsection{Database and \feature\ Extraction}%
	\label{sec:data}

We experiment with non-parallel bimodal data (\EEG\ and Speech) for multi-modal emotion analysis because of the absence of 
publicly available \textit{parallel dataset} having Speech and \EEG\ 
modalities 
recorded %
for
the same stimuli, simultaneously. %
The unavailability 
can be attributed to the %
contradictory recording condition required for \EEG\ and Speech, %
while 
the former requires a subject's head to be %
as still as possible, while the %
latter requires a %
 significant jaw movement to speak which 
 would in turn induce noise in 
 \EEG, if both 
 modalities were 
 recorded in parallel. %

For \EEG, we 
 use the 
 \DEAP\ dataset   \cite{koelstra2011deap} consisting of \EEG\ data with %
 valence (\Val), arousal (\Aro), dominance (\Dom) and liking (\Lik)
	emotion %
 labels. 
	\DEAP\ dataset elicits emotion in response to a $1$ minute %
 audio-visual stimuli.
 There are $32$ subjects, each %
 shown $40$ clips (trials) during which %
 $7$ physiological modalities were recorded. Each trial has physiological recording of $40$ channel in which first $32$ channels corresponds to \EEG. Additionally a %
 self rated emotion %
 score (between $1$ and $9$) for each %
 trial is available.  %
 In our experiments %
 we %
 drop the $3$s pre-trial data %
 and
 use only $60$s of the $63$s data per trial. %
 Furthermore, we %
 follow 
 the  pre-processing %
 adopted in \cite{koelstra2011deap}, 
 to construct
 binary labels \high\ ($\ge \lambda$) and \low\ ($<\lambda$) on the self-assessment labels with  $\lambda = 4.5$
Thus %
each trial, had one of two labels in the  \Val-\Aro-\Dom-\Lik\ space, namely,
 \Val\high\  or \Val\low; \Aro\high\ or \Aro\low; \Dom\high\ or \Dom\low; and \Lik\high\ or \Lik\low.
For proposed \jmml\ approach, we make use of \Val-\Aro\ emotion space of the \DEAP\ EEG data. %
 As mentioned earlier, %
 we extract nine tempo-spectral features for 
 each of the 
 five standard EEG bands %
 using \texttt{PyEEG} \cite{bao2011pyeeg} at the trial-level without any frame-level computation. 
Furthermore, we perform \feature\ selection to reduce the high dimension and information redundancy resulted from 32 \EEG\ channels. 
We evaluated all possible subset combination of EEG bands and features to select best features and bands.
The final selected features are Hjorth features, HFD, PFD,
Spectral Entropy, PSI, RIR 
in the 
$\alpha_{low}$, $\alpha_{high}$, $\beta$, $\gamma$ bands. This resulted 
in a
416 dimensional EEG feature vector.
	
For speech data, we use fusion of two standard audio emotion datasets, namely, (A) Berlin Emotional Database ({\sc Emo-DB}) \cite{EmoDB} which consists of 535 acted utterances recorded by rehearsing the memorised script in fairly clean environment, eliciting
7 emotion categories and, (B) the Ryerson Audio-Visual Database of Emotional Speech and Song ({\sc RAVDESS}) \cite{livingstone2018ryerson} which consists of 1440 samples, where the participants
 vocalized lexically-matched statements in a neutral North American accent, eliciting
7 emotion categories. In our experiments, we reduce the multi-class speech data to binary class by considering 
the most misclassified emotion class pairs, i.e, {\tt Anger}-{\tt Happy}, and {\tt Neutral}-{\tt Sad}.
Lastly, in order to jointly train the \EEG\  and speech samples, the categorical emotion label of audio samples are converted into dimensional label to match with \EEG\  labels. 
This is done by relabelling {\tt Anger} as \Val\low, {\tt Happy} as \Val\high, {\tt Sad} as \Aro\low\ and {\tt Neutral} as \Aro\high.
We extract, 
as mentioned earlier a
$88$ dimensional acoustic feature vector %
for every audio sample.

\subsection{Model Configuration and Training}
\label{models}
Firstly, we perform \jec-\ssl\ (Figure \ref{fig:jec-ssl}) %
for each modality, namely, \EEG\  and speech, with binary  (\high, \low) class labels.
Thereafter, the output of the \jec-\ssl\ is fed as input to perform the \edcc-\cae\ (Figure \ref{fig:jmml}) based \jmml.

\subsubsection{\jecl}
\label{sec:exp_jecl}
We implement \jecl\ with two emotion blocks ($\B_1$ and $\B_2$) as shown in Figure \ref{fig:jec-ssl}. 
\AE\ %
for both \textit{independent} and \textit{similarity} branch %
consists of $3$ layers with {\em ReLU} activation, namely, \E, \L\ and \D.
To keep the \AE\ compact we stacked only single \E, \L\ and \D\ layers in
each branch. We tried three different setups for selecting the number of hidden neurons in each layer;
  same number of neurons in \E, \L\ and \D\ (\setup{1}), compressed latent space (\setup{2}) with
number of hidden neurons in \L\  as half of that in \E\ and \D, and   expanded latent space (\setup{3}) with
number of hidden neurons in \L\ as twice of that in \E\ and \D. Each of the above mentioned setups are
tried with $N/2$, $N$, $2N$, $4N$ number of hidden neurons, where $N$ is 
the
dimension of the input vector. We found \setup{3} with $2N$ neurons in \E\ and \D\ and $4N$ in \L\ to be working best.
Further, the decoded output from both the branches are concatenated and fed to the final fully connected ({\sc fc}) dense layer with {\em linear} activation and $N$ neural units. %
 The final output of each \jecl\ block is the joint representation learnt per emotion class. So, we get %
 one $N$-dimension
 \jecl\ output vectors %
 for each block. %
 The latent layer $\L_{sim}$ of the similarity branches from two blocks are tied together (by sharing weights) unlike the $\L_{ind}$  (independent) branches. %
{\em We hypothesize that this process of joint %
 learning %
 helps %
 capture not only 
 class-specific emotion properties but also similarities across %
 different
  emotion classes.}

	As an example, %
 assume $\vX_{\train}$ be the training set that consists of two class data $\vX_{\train}^1$ and $\vX_{\train}^2$. During training, %
 $\vX_{\train}^1$ is input to block $\B_1$ (in an epoch) while $\vX_{\train}^2$ is input to $\B_2$ in a sequence. %
	At each epoch $e$, %
 both $\B_1$ and $\B_2$ are trained. %
 While the shared latent space weights ($\L_{sim}$) are updated 
 for every epoch, irrespective of the block the  %
 block layer weights ($\L_{ind}$, $\E_{ind}$, $\D_{ind}$, $\E_{sim}$, $\D_{sim}$) are updated only once per epoch when that block sees an input.
	The \jecl\ is %
 implemented in \texttt{Keras} \cite{Keras} and are trained using the \texttt{adam} optimizer with customised loss %
 (as discussed in section \ref{sec:jecl}) for a number of epochs %
 guided by the validation loss.

 We use   \texttt{mbpls} python package, to implement the \mbpls\ model with two data blocks %
 consisting of $N$-dimensional joint embeddings learnt from both $\B_1$ and $\B_2$ of \jecl.
	Note that %
 \mbpls\ is trained on \jecl\ output %
 and maps them to a common latent subspace. The target vector of \mbpls\ is the original train data itself. 
	From these two data blocks, 
	 \mbpls\ predicts a $N$-dimensional vector, such that respective contribution of each emotion block is retained. 
This vector is the final joint emotion class embedding which is used for %
emotion classification. 
We tune \mbpls\ $LVs$ using range of $LVs$ starting from 40 to 120 with an increment of 2.

\subsubsection{\edcc-\cae\ based \jmml}

We implement \jmml\ for a bimodal cross-corpus (/non-parallel) data, 
\EEG\  data represented as $\vX_1$ and %
speech data represented as $\vX_2$ as shown in Figure \ref{fig:jmml}. 
\jmml\ is implemented using a \AE\ branch which is stacked with a DNN based encoder $\mbox{DNN}$-$E_m$ followed by a DNN based decoder $\mbox{DNN}$-$D_m$, for each modality $m$. 
Each encoder ($\mbox{DNN}$-$E_1$ and $\mbox{DNN}$-$E_2$) and decoder ($\mbox{DNN}$-$D_1$ and $\mbox{DNN}$-$D_2$)
consists of $3$ hidden layers and an output layer. 
Each hidden layer 
is trained with {\em ReLU} activation. 
To tune the hidden units, we adopted the similar approach as discussed in Section \ref{sec:exp_jecl} for \jecl\ training.
The output layer of each encoders (${\vX'}^{oe}_1$ and ${\vX'}^{oe}_2$) has $20$ units with linear activation function which is merged and correlated with $\mbox{\CCA}$ loss \cite{andrew2013deep}. We use {\tt keras} with {\tt tensorflow} as %
backend to implement this loss. 
The output of penultimate layer of each encoders (${\vX'}^{e}_1$ and ${\vX'}^{e}_2$) is further used by the decoders %
for the reconstruction. 
Lastly, each decoder has two output nodes with linear activation, for \crossreconstruction\  and \selfreconstruction\, respectively. For \EEG, the reconstructed output $\bar \vX_1^2$ is the \crossreconstruction\
speech representation mapped from \EEG. Similarly, the reconstructed output $\bar \vX_2^1$ is the \crossreconstruction\
\EEG\  representation mapped from speech. Besides learning the cross mapping, each modality specific decoder also outputs it's own mapping represented by \selfreconstruction, namely, $\bar \vX_1^1$ and $\bar \vX_2^2$.
The overall \jmml\ is trained using {\tt Adam} optimizer with a batch size of $32$
with an overall loss (Equation \ref{eq:my_jmml_loss})
for a number of epochs guided by the validation loss.

\section{Experimental Results and Analysis}
\label{sec:results}

We split the data into %
train set ($80\%$) and  test set ($20\%$) for both the EEG and speech data. 
Further $10\%$ of the train data is used for validation. %
The EEG data consisted of (\Val+: 808, \Val-: 472; \Aro+: 818, \Aro-: 468) samples while the speech data has (\Val+: 210, \Val-: 255; \Aro+: 140, \Aro-: 203) samples. 
For EEG data, we adopted Minority Class Oversampling (MCO) to overcome the %
class imbalance %
across %
the \Val-\Aro\
emotion dimensions.
 Additionally, 
 we applied the same oversampling technique to the speech data 
 so that the number of speech samples matched the samples in EEG.
 We build %
 a 
 baseline emotion recognition system using the features mentioned in Section \ref{sec:data} and Random Forest (\RF) as the final stage classifier, for both EEG and speech %
 modalities. 

\begin{table*}[tb]
\centering
\caption{Baseline vs \jec-\ssl\ and \jmml\ for both Valence and Arousal Emotion Recognition.}
\label{tab:JMML-VA}
\resizebox{0.8\textwidth}{!}{ 
\begin{tabular}{|c|ccc||ccc|}\hline
\multicolumn{7}{|c|}{Valence}\\ \hline
\multirow{2}{*}{Experiment Setup} & \multicolumn{3}{c||}{\EEG}& \multicolumn{3}{c|}{Speech}\\ \cline{2-7} 
& \multicolumn{1}{c|}{Input} & \multicolumn{1}{c|}{Acc} & F1 & \multicolumn{1}{c|}{Input} & \multicolumn{1}{c|}{Acc}& F1\\ \hline

Baseline & \multicolumn{1}{c|}{$\vX_1$} & \multicolumn{1}{c|}{70.4} & 69 & \multicolumn{1}{c|}{$\vX_2$}& \multicolumn{1}{c|}{78.6} & 78.77 \\ \hline
\jec-\ssl\                          & \multicolumn{1}{c|}{$\vX'_1$}      & \multicolumn{1}{c|}{72.4} & 71.5 (+2.5\%) & \multicolumn{1}{c|}{$\vX'_2$}      & \multicolumn{1}{c|}{82.9} & 82.9 (+4.2\%)\\ \hline\hline
Baseline $\rightarrow$ \edcc-\cae     & \multicolumn{1}{c|}{{[}$\vX_1$,$\bar {\vX}^1_1${]}}      & \multicolumn{1}{c|}{71.6}  & 70.2 (+1.2\%)& \multicolumn{1}{c|}{{[}$\vX_2$,$\bar {\vX}^2_2${]}}      & \multicolumn{1}{c|}{82.1} & 82.1 (+3.4\%)\\ \hline
\begin{tabular}[c]{@{}c@{}}\jmml\ [\jec-\ssl\ $\rightarrow{}$ \edcc-\cae]\\(\textbf{Proposed approach})\end{tabular}     & \multicolumn{1}{c|}{{[}$\vX'_1$, $\bar {\vX'}^1_1${]}}      & \multicolumn{1}{c|}{72.8}   & 71.8 (+2.8\%)& \multicolumn{1}{c|}{{[}$\vX'_2$, $\bar {\vX'}^2_2${]}}      & \multicolumn{1}{c|}{85.6} & 85.6 (+6.9\%)\\ \hline\hline
\end{tabular}
}
\centering
\resizebox{0.8\textwidth}{!}{ 
\begin{tabular}{|c|ccc||ccc|}\hline
\multicolumn{7}{|c|}{Arousal}\\ \hline
\multirow{2}{*}{Experiment Setup} & \multicolumn{3}{c||}{\EEG} & \multicolumn{3}{c|}{Speech}  \\ \cline{2-7} 
& \multicolumn{1}{c|}{Input} & \multicolumn{1}{c|}{Acc} & F1 & \multicolumn{1}{c|}{Input} & \multicolumn{1}{c|}{Acc}   & F1    \\ \hline
Baseline                          & \multicolumn{1}{c|}{$\vX_1$}    & \multicolumn{1}{c|}{69.5} & 68.5 & \multicolumn{1}{c|}{$\vX_2$}    & \multicolumn{1}{c|}{81.4} & 81 \\ \hline

\jec-\ssl\                          & \multicolumn{1}{c|}{$\vX'_1$}      & \multicolumn{1}{c|}{72.3} & 72.6 (+4.1\%)& \multicolumn{1}{c|}{$\vX'_2$}      & \multicolumn{1}{c|}{83.7} & 83.8 (+2.8\%) \\ \hline\hline

Baseline $\rightarrow$ \edcc-\cae      & \multicolumn{1}{c|}{{[}$\vX_1$,$\bar {\vX}^1_1${]}}      & \multicolumn{1}{c|}{71.5}  & 70.4 (+1.9\%)& \multicolumn{1}{c|}{{[}$\vX_2$,$\bar {\vX}^2_2${]}}      & \multicolumn{1}{c|}{84.9} & 84.5 (+3.5\%) \\ \hline

\begin{tabular}[c]{@{}c@{}}\jmml\ [\jec-\ssl\ $\rightarrow{}$ \edcc-\cae]\\(\textbf{Proposed approach})\end{tabular}      & \multicolumn{1}{c|}{{[}$\vX'_1$, $\bar {\vX'}^1_1${]}}      & \multicolumn{1}{c|}{71.9}   & 70.6  (+2.1\%) & \multicolumn{1}{c|}{{[}$\vX'_2$, $\bar {\vX'}^2_2${]}}      & \multicolumn{1}{c|}{88.4} & 88.2 (+7.2\%)\\ \hline
\end{tabular}
}
\end{table*}

We evaluate the proposed \jmml\ approach by performing the final stage emotion recognition in four different experimental setups as shown in Table \ref{tab:JMML-VA} %
for both \Val\ and \Aro\ emotions. %
We use 
\RF\ as the  final stage classifier %
in %
all our experiments. %
We perform grid search to fix \RF\ parameters \textit{n\_estimators} and \textit{n\_depth} for each of our experimental setup independently, with the grid of $n\_estimators= (i*10)$, where $100 \le i \le 500$, and $n\_depth= (2*i)$, where $1 \le i \le 20$. We discuss each experimental setup in detail as below.

\begin{enumerate}
    \item{Baseline:} Firstly, we create an individual baseline for both the modalities using the $416$-D tempo-spectral feature vector ($\vX_1$) for \EEG\  and $88$-D acoustic feature vector ($\vX_2$) for speech (as discussed in Section \ref{sec:features}). 
    \def\orig#1#2{$#1_{#2}$}
    \def\srec#1#2{$\bar #1^{#2}_{#2}$}

    \item{\jec-\ssl:} This setup performs only intra-modal learning using \jec-\ssl (as discussed in Section \ref{sec:jecl}) with original representation ($\vX_1$ for \EEG\ and $\vX_2$ for Speech) as input. 
    \jec-\ssl\ shows an absolute improvement over Baseline for both EEG and Speech modality in terms of both accuracy and F1 scores. As seen in Table \ref{tab:JMML-VA}, for EEG modality there is an absolute improvement of $+2.5\%$ and $+4.10\%$ for \Val\ and \Aro\ respectively in terms of the F1 score, while the F1 score improvement for speech modality is $+4.2\%$ for \Val\ and  $+2.8\%$ for \Aro.

    \item {Baseline $\rightarrow$ \edcc-\cae:} We perform this experiment to evaluate the impact of inter-modal information on AER performance,
    explicitly, independent of intra-modal learning. 
    The input feature vector $\vX_m$ for $m=[1,2]$ is fed directly to \edcc-\cae\ module with no intra-modal learning (\jec-\ssl) in between. We evaluate this setup performance against the Baseline system. The original \EEG\ representation ($\vX_1$) and Speech representation ($\vX_2$) is fed to the %
    input node of each modality, respectively. It is to be noted that the proposed \edcc-\cae\ system (as discussed in Section \ref{sec:jmml}) allows the independent inference of either modalities, i.e, \EEG\ data is not required to be present when using Speech as the modality during %
    inference, and vice-versa.
    We %
    use \selfreconstruction\ output vector for final evaluation in the both modalities. %
    We concatenate \selfreconstruction\ output (\srec{\vX}{1} for \EEG\ %
    and \srec{\vX}{2} for speech) %
     with 
     original input representation (\orig{\vX}{1} for \EEG\ and \orig{\vX}{2} for speech) %
     to perform the final evaluation. %
    Concatenating the \SREC\ %
    with the original input representation, namely 
    $[\vX_m,\bar {\vX}^m_m]$, %
    results in a significant improvement in emotion classification over baseline. 
    Specifically, an improvement  of $+1.2\%$ and $+2\%$ is observed in terms of F1 score for \Val\ and \Aro\ respectively as seen in Table \ref{tab:JMML-VA} for EEG modality. For the speech modality the improvement in terms of F1 score for \Val\ and \Aro\ is $+3.4\%$ and $+3.5\%$ respectively.

     \item{\jmml [\jec-\ssl\ $\rightarrow$ \edcc-\cae]:} Finally, we report the performance of the proposed two-step \jmml\ approach, namely, %
     joint intra-modal learning (\jec-\ssl) followed by joint inter-modal learning (\edcc-\cae). 
     Similar to Baseline $\rightarrow$ \edcc-\cae\ experimental setup, we concatenate the \selfreconstruction\ output (\srec{\vX'}{1} for \EEG\ modality; \srec{\vX'}{2} for speech modality) %
     with \jec-\ssl\ based input representation (\orig{\vX'}{1} for \EEG\ and \orig{\vX'}{2} for speech modality) to perform the final evaluation. 
     The \jmml\ results in a considerable
     improvement in F1 score for speech modality for both \Val\ $(+6.9\%)$ and \Aro\  $(+7.2\%)$ as seen in Table \ref{tab:JMML-VA} 
      over %
    the Baseline.

\end{enumerate}

 It is to be noted that the SER performance using \jmml\ %
 for speech modality %
 not only surpasses the Baseline, but it also shows %
 a significant improvement over \jec-\ssl. Specifically, \jmml\ betters \jec-\ssl\ by  $+2.7\% = (85.6 - 82.9)$ for \Val\ and $+4.4\% =(88.2-83.8)$ in terms of F1  score as seen in Table \ref{tab:JMML-VA}. 
 However, %
 for \EEG\ modality, the proposed \jmml\ approach provides a very moderate performance improvement over \jec-\ssl. 
This minimal performance in the EEG modality can be attributed to 
performance saturation, where \jec-\ssl\ has already achieved SOTA performance \cite{tiwari2024joint}, leaving it less room %
to improve further. %
The significant improvement (over \jec-\ssl) in SER performance using \jmml, clearly demonstrates the efficacy of the proposed approach.
As can be seen, combining the intra-modal (\jec-\ssl) and inter-modal characteristics (\edcc-\cae) through the proposed \jmml\ approach, %
shows that Speech modality benefits from 
the 
EEG data during \jmml\ learning
resulting in improved SER performance.
Note that we make no effort to compare our results with \cite{liu2024eeg}, because
unlike \cite{liu2024eeg} 
which %
fuses EEG and Speech modalities either at the feature-level or at the decision-level %
to perform multi-modal AER  we adopt a novel attempt
where the Speech modality is facilitated with the EEG modality by coordinating both modalities via joint multi-modal learning and there is no attempt to fuse information.

\section{Conclusion}
\label{sec:conclu}
This paper introduces
a novel two-step joint multi-modal learning (\jmml) %
framework,
where joint-class intra-model learning is followed by an 
inter-modal 
joint learning %
for improved and reliable 
performance of 
automatic emotion recognition. %
Our work, 
specifically 
targets to facilitate
the performance of emotion recognition on speech modality, which is
the most popular and easy-to-use modality but unreliable, %
with 
the help of EEG,
one of the most reliable modality %
used to 
identify 
human emotion.
It should be noted that, though reliable, using 
EEG is challenging %
due to it's complex recording requirement during practical deployment.
Our approach of %
joint inter-modal learning addresses this challenge by exploiting 
the correlations between 
EEG and Speech  modalities
in such a way that the 
non-parallel 
bi-modal (i.e EEG and Speech together) data is required only during 
model
training
and not during inference.

During the two step training, %
the process of 
joint intra-modal learning (Step 1),  %
captures, within each modality,
(a) the common characteristics %
across
all the emotion classes %
in addition to 
the %
characteristics
very specific to a given emotion class.
This 
is %
followed by the joint inter-modal learning (Step 2) using 
the proposed 
\edcc-\cae. 
We eliminate the need for prior knowledge of the missing modality
during
inference by enabling %
the joint inter-modal learning %
to not only self-reconstruct (\selfreconstruction) itself but
also cross-reconstruct (\crossreconstruction) the other modality. 
In a nutshell,
the joint inter-modal learning (using \edcc-\cae) makes use of complementary \EEG\ %
modality
to enhance the Speech representation %
during joint training of both modalities, 
while keeping the modalities %
independent of each other
at the time of inference. %
This way, the emotions can be reliably detected at the time of inference
from even a single modality. 
Specifically, %
it does not effect the performance of the emotion recognition using the speech modality even in the absence of  a %
reliable %
modality like EEG %
during inference. %
This enables building usable emotion recognition 
systems %
in 
realistic scenarios where it is difficult to %
capture 
EEG data at the time of
inference.

\bibliographystyle{cas-model2-names}
	\bibliography{mybib}

\end{document}